\documentclass[sn-apa]{sn-jnl}% APA Reference Style 
\usepackage{graphicx}%
\usepackage{multirow}%
\usepackage{amsmath,amssymb,amsfonts,bm}%
\usepackage{physics}%
\usepackage{amsthm}%
\usepackage{mathrsfs}%
\usepackage[title]{appendix}%
\usepackage{xcolor}%
\usepackage{textcomp}%
\usepackage{manyfoot}%
\usepackage{booktabs}%
\usepackage{algorithm}%
\usepackage{algorithmicx}%
\usepackage{algpseudocode}%
\usepackage{subcaption}
\usepackage{listings}%
\usepackage{csquotes}
\usepackage{hyperref}
\algnewcommand{\LineComment}[1]{\State \(\triangleright\) #1}
\usepackage[colorinlistoftodos]{todonotes}

\newlength{\tempdima}
\newcommand{\rowname}[1]% #1 = text
{\rotatebox{90}{\makebox[\tempdima][c]{\textbf{#1}}}}

\begin{document}

\title[Article Title]{Benchmarking the ORCA PT-2 Boson Sampler using Minimum Dominating Set Problems}


\author*[1,2]{\fnm{Jessica} \sur{Park}}\email{jlp567@york.ac.uk}

\author[1]{\fnm{Susan} \sur{Stepney}}\email{susan.stepney@york.ac.uk}
\equalcont{These authors contributed equally to this work.}

\author[2]{\fnm{Irene} \sur{D'Amico}}\email{irene.damico@york.ac.uk}
\equalcont{These authors contributed equally to this work.}

\affil*[1]{\orgdiv{Department of Computer Science}, \orgname{University of York}, \orgaddress{\city{York}, \country{UK}}}

\affil[2]{\orgdiv{Department of Physics}, \orgname{University of York}, \orgaddress{\city{York}, \country{UK}}}

\abstract{We use boson sampling as part of a gradient-free variational algorithm (the Binary Bosonic Solver) to solve a minimum dominating set problem and compare these results to a number of exact and heuristic classical algorithms. The boson sampling has been performed on the physical PT-2 time-bin interferometer from ORCA Computing. The PT-2 device has been tested here using both a single- and double-loop configuration and the results are compared based on the best found solution and the overall run time. With the parameters used in this experiment, the boson sampler is outperformed by the classical methods, but we hypothesise that this is due to insufficient samples and iterations. We classically simulate boson sampling in a single-loop configuration to break down the runtime for individual algorithmic components, allowing for estimates of when boson sampling may outperform classical methods. This study recommends a watching brief on boson sampling as the complexity of the interferometer is improved and the loss in the hardware is reduced allowing for better performance from the associated algorithms.}

\keywords{Optimisation, Boson Sampling, Benchmarking}



\maketitle

\section{Introduction}\label{sec1}
Boson sampling is a quantum computing paradigm that is based on determining the output distribution from bosons in linear interferometers based on knowledge of the beam splitter locations and angles, the number of input bosons, and the number of output modes. 
Classically, this is computationally very expensive to calculate (\#P complete, so at least as hard as NP-complete) with an exponential number of terms. However, a boson sampling device enacts the chosen interferometer and reports the output \citep{Aaronson2011-tm}. 
Although a boson sampler can produce this output distribution more effectively than a classical computer could calculate it, there is ongoing debate as to whether a boson sampler can solve any real problems with a quantum advantage \citep{Bromley2020-uc}. 
It has been shown that boson samplers can be used in variational algorithms (eg. combinatorial optimisation, generative adversarial networks), calculating graph properties and proof-of-work consensus, although they have yet to overtake the utility of classical supercomputers \citep{Slysz2024-hs, Bacarreza2025-xu, Arrazola2018-gp}. 

Our previous work  \citep{Park2026-sy} uses simulation to show that a boson sampler can, in principle, be used to find solutions to dominating set problems that are of comparable quality to typical classical methods. 
That work uses the ORCA Computing Software Development Kit (SDK) with their Binary Bosonic Solver (BBS) to simulate both their PT-1 and PT-2 devices.
The classical methods used for comparison were a greedy algorithm, a heuristic algorithm from the NetworkX Python package, and a linear programming approach using the PULP Python package.
The wall-clock time required to run the boson sampler simulation is orders of magnitude larger than the classical approaches, but this is to be expected from the complexity of calculating a linear interferometer output distribution as described above. 

We extend the results presented previously in this work,
by replicating the benchmarking experiment using a physical ORCA PT-2 series boson sampler, rather than its simulator. 
The PT-1 is no longer in use, and therefore cannot be used for comparison. 
There have also been a number of significant updates in the BBS algorithm, designed to improve performance and runtime. 
Therefore this study reports new simulated results as well as the results from the real device, to ensure a fair comparison. 
The classical algorithm results from the original paper are repeated here for comparison purposes.

\section{Background}
\subsection{Photonic boson sampling}
%\notejp{This theory section is pretty light touch compared to the original UCNC paper but I think it still works as a standalone. This might depend on where this ends up going as a computer science audience might need more details on interferometry.}
%\notess{Do you have a reason not to include the original UCNC text?  It's allowed for an extension paper (although not for a stand-alone).}

A photonic boson sampler, such as the ORCA PT-2, performs calculations using samples from a distribution of identical photons after they have passed through a linear interferometer. 
Linear interferometers are formed of a number of optical components called beam splitters that can change the path of incident photons. 
Changing the `angle' of the beam splitter ($\theta$) changes the probabilities of transmission or reflection of the photon.
Quantum effects, namely the Hong-Ou-Mandel effect, appear when multiple indistinguishable photons are coincident on a beam splitter.
The Hong-Ou-Mandel effect describes the quantum interference of indistinguishable photons at a beam splitter whereby the photons will always leave the beam splitter in the same output mode \citep{Hong1987-jq}.
They are then in a superposition of both output paths and the result is determined on measurement.

Photons are commonly used in boson samplers as they have long coherence times, do not interact strongly with the environment, and can function at room temperature. 
Interferometers can be constructed either in free space or with the photons travelling through optical fibres between beam splitters.
Because they travel at the speed of light, the construction of the linear interferometer needs to be very precise to ensure that photons arrive at the beam splitters at the same time and are indistinguishable, allowing for the quantum effects to take place. 

Linear interferometers are usually diagrammatically and physically spatially binned. 
This means that the input photons all enter the interferometer at the same time in different channels that are directed towards a rows of beam splitters. 
The width of the interferometer refers to how many channels the photons can enter the interferometer by and the depth refers to the number of rows of beam splitters through which the photons will pass.
At the end of each channel in the interferometer, there are photon detectors which then measure the output distribution. 
The number of output bins is typically known as the number of modes of the interferometer.

\subsection{Physical devices}

\begin{figure}
    \centering
    \includegraphics[width=0.95\linewidth]{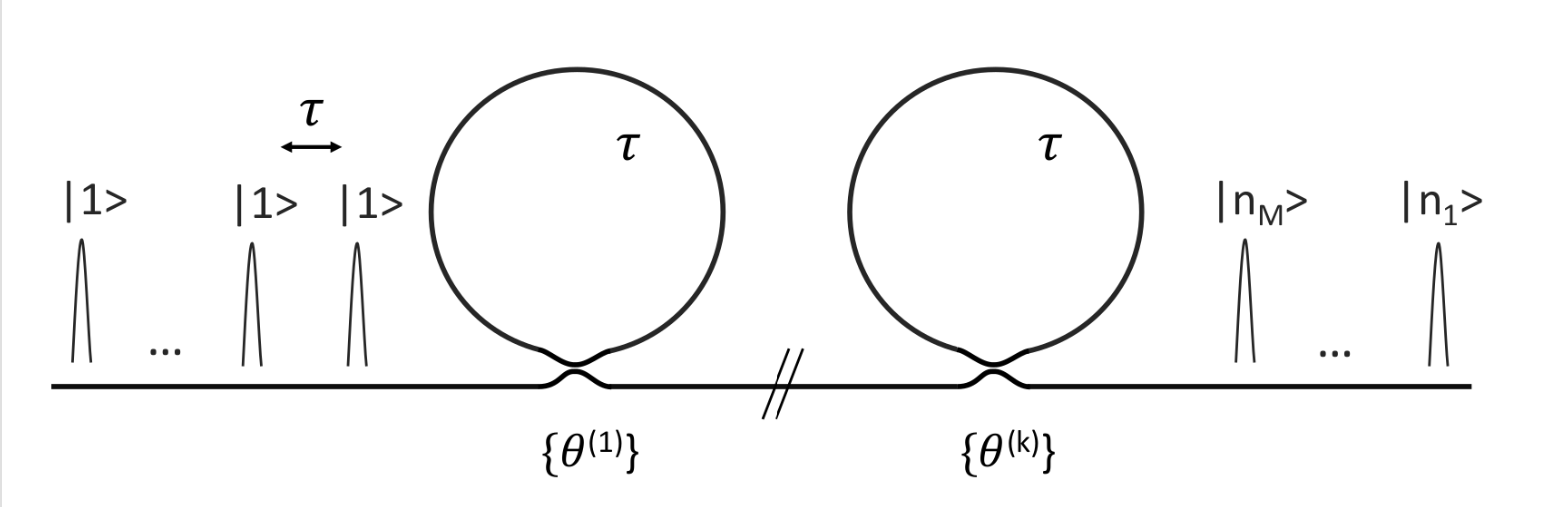}
    \caption{Diagrammatic representation of time-bin interferometry. Taken from  \citet{Bradler2021-xa}.}
    \label{fig:tbi}
\end{figure}

ORCA Computing, among others, have a different approach to the interferometer that requires less physical space and fewer components.
They design and build time-binned devices, in which the input photons enter the interferometer via a single channels at defined intervals, $\tau$, and loops of fibre-optic cable are used as delay lines to cause photons to interact at the beam splitters \citep{Bradler2021-xa}.  
At each beam splitter, the photon can either move through to the next stage of the interferometer or be directed back into a fibre optic loop where it could then interact with the following photon mode at the same beam splitter. 
This means fewer beam-splitters are required, and only a single photon generation device. 
Similarly, this set-up only requires one photon detector, as the output modes, $\ket{n_M}$ are counted as the time steps in which photons exit the interferometer and are detected \citep{Motes2014-xj, He2017-pc}.
The principle of time-bin interferometry is shown in Figure~\ref{fig:tbi} and is mathematically equivalent to the spatial approach \citep{Sempere-Llagostera2022-wz}.

Our previous work  \citep{Park2026-sy} mainly considers the ORCA PT-1 device, as that was the available device at the time.
The PT-2 device is mentioned in that study as a future development planned by ORCA Computing.
Since then ORCA Computing have begun manufacturing, selling and operating PT-2 devices. 
The PT-2 device samples from non-Gaussian, photon-subtracted squeezed states, rather than single photons as in the PT-1, and uses a photon-number resolving detector \citep{Olson2015-zj}. 
There have also been engineering improvements for more stability and lower loss. 
The PT-2 device also has capability for running a time-bin interferometer with two loops. 
By analogy to the spatial interferometer, adding more loops provides additional depth or layers of beam splitters, whilst inputting more photons (more time-bins) is the equivalent to making the interferometer wider and having more spatial modes.
%\notess{Can you quickly explain how a second loop gives more computation power/ability?  Is it just like adding an extra beam-splitter, or does it allow something else/more?- Added the extra sentence and some references}
Each additional loop makes the PT Series progressively harder to simulate, with only 3 loops required to reach a regime that is not simulable classically \citep{Novak2025-ew, Deshpande2022-lu}. 
Whether this theoretical quantum advantage translates to real world quantum utility in the kind of applications considered here remain unproven, and will depend on whether generating more complex output distributions brings an advantage in training optimisation algorithms.

\subsection{Optimisation algorithm}

The Binary Bosonic Solver (BBS) is a variational quantum algorithm that uses the output distribution from the time-bin interferometer to solve optimisation functions. 
It is an iterative approach involving the training of beam splitter parameters to optimise a given cost function. More details are available in \citet{Makarovskiy2025-kl} and its application in this study is detailed in the following section.

The key development since our previous work is the option for a gradient-free version of the BBS. 
This requires evaluating a single observable per update step instead of multiple evaluations per update step to calculate gradients.  
It is expected that the repeated cost function evaluations make up a considerable portion of the total algorithm runtime; if this is so, fewer evaluations should improve runtime. 
The timings for the individual portions of the algorithm are discussed alongside the results below.

\subsection{Application domain}

Due to the significant upgrade in both the hardware and software since our original study, we do not provide a direct comparison between those results and the new simulations here. 
Instead, we run further simulations using the new gradient-free binary bosonic solver.

As in the original study, we are considering the use of boson sampler in the case of solving a surveillance coverage problem that can be articulated as follows.
Given a map of places of interest (POIs), a graph can be drawn where each POI is a node and an edge exists between nodes where there is a line of sight between them. The surveillance coverage problem is to find the smallest set of POIs that have the ability to surveil the entire map.
We use here the convention that a node within the surveillance set can surveil itself.
Mathematically, this is equivalent to the NP-hard \textit{minimum dominating set} problem \citep{Grandoni2006-mz}:
for a graph $G=(V,E)$, find the smallest subset $V'$ of $V$ such that every node in the graph is adjacent to at least one node in  $V'$.

\section{Experimental Design}\label{sec:meth}
The majority of the experimental design follows that laid out in our earlier work \citep{Park2026-sy}, in which the binary bosonic solver (BBS, from ORCA Computing) is applied to dominating set problems of increasing size. 
As discussed in the previous section, the BBS algorithm has been updated by ORCA Computing to use gradient-free optimisation.
This updated version is used in this study.
%\notess{Is the BBS the same as the CPU-based sim of PT-1? - Updated text

%\notejp{edited below} The algorithm is executed using both the ORCA PT-2 hardware in both a 1-loop and 2-loop configuration, and as a CPU-based 1-loop simulation of the ORCA device.
These results are compared against three classical alternatives: a greedy algorithm, the NetworkX approximation algorithm, and an exact linear programming approach. 
These six experiments are laid out in Table~\ref{tab:expts}. 
All of the experiments are run on identical test problems to enable accurate comparison. 
The CPU used to produce the classical results in this study runs an Intel i7 processor with 64GB of RAM. 
That methodology is summarised here with the key new developments and changes highlighted.
%\notess{This appears to be the old text -- what of the PT-2?
% It might be useful to have a little table of all the expts,
% laying out what is done in hw, what in sim, what classically (eg, something like the key in fig 3, but with some explanation) - }

\begin{table}
    \centering
    \begin{tabular}{ccc}
        \toprule
        \textbf{Algorithm} & \textbf{Hardware} & \textbf{Label}\\
        \midrule
        Binary Bosonic Solver & PT-2 (1-loop Implementation) & BBS 1-loop PT-2\\
        Binary Bosonic Solver & PT-2 (2-loop Implementation) & BBS 2-loop PT-2\\
        Binary Bosonic Solver & CPU (1-loop Simulation) & BBS 1-loop Sim\\
        Greedy & CPU & Greedy \\
        NetworkX Approximation & CPU & NX Approximation \\
        Exact Linear Programming & CPU & PULP\\
        \bottomrule
    \end{tabular}
    \caption{Table showing the six experiments run and compared during this study. The label column corresponds to the legends across all figures in the results section.}
    \label{tab:expts}
\end{table}

\subsection{Dominating set cost function}
The particular graphs that the algorithms are run on are the same as in the original work. They are defined as  $G(n,p,s)$, where $G$ is a graph with $n$ nodes and a $p$ probability of edge creation between any two nodes; $s$ defines the random seed that is used to ensure repeatable graphs are produced.
For all the results presented here we fix the graph probability density as $p=0.05$.
For meaningful results, the graph needs to be connected; this is ensured in the graph generation process. 
For each $n$, three random graphs are generated by using three different values of the seed $s$. 
This set of graphs is fixed for use with both the quantum algorithm and classical comparison. 

We formulate a dominating set problem as follows.
Let $\bm{x}$ be the bit string $\bm{x} =[x_0, x_1, ..., x_{n-1}]$, where $n$ is the number of vertices in $G$ and $x_i = 1$ if the vertex is in the candidate set and $0$ otherwise. The neighbours of the vertex $x_i$ are $N(x_i)$.
The cost function to be minimised to find the minimum dominating set is:

\begin{equation} \label{eq:F}
    F(\bm{x}) = \sum_{v_i \in V}x_i + A\sum_{v_i \in V}P_i,
\end{equation}
where $P_i$ is the penalty term given by
\begin{equation} \label{eq:P}
    P_i = 
    \begin{cases}
    0,& \text{if } (x_i + \sum_{v_j \in N(v_i)}x_j) -1 \ge 0\\
    1,& \text{otherwise}
    \end{cases}
\end{equation}

The first term in equation~\ref{eq:F} is the size of the candidate set.
The second term is the penalty term, where $A$ is a scaling factor (in this case, $A=2$) and $P_i$ is a function that checks whether a vertex $v_i$ is either part of, or neighbours a vertex in, the set. 
A penalty of $1$ is added for every vertex in the graph that is not either part of the set or neighbours with at least one node that is. 
If no penalty terms are added, the set in question is a dominating set, and the result of the cost function is the size of the set.
We refer to the result of the cost function, $F(x)$, as the state \textit{energy} of that bit string. 

For some problems, a minimal dominating set is not a unique solution. It is relatively easy to check whether a set of nodes dominates the graph, but without exhaustive search, it is impossible to know if it is a \textit{minimal} dominating set.
The output from each of the different algorithms is a set of nodes which is then verified to be dominating (a computationally easy step).
Without knowing the true minimum set size, we are comparing the found set size from the different methods where a smaller set implies a more performant algorithm. 
%\notess{So we are solving only the smallish dominating set problem?  Need to make this clear -- the output is a dominating set (is this actually checked?) but not nec. the minimum one -- so will be comparing performance both in time/efficiency, and in smallness of result? - Clarified}
%Here we work with the common variant in which a node can dominate itself. 
%When this is not the case, and a node cannot dominate itself, it is known as a strongly-dominating set or a total dominating set.
%The only change in the cost function to enact that problem is to remove the $x_i$ from the the first statement in equation~\ref{eq:P}. That then requires that the vertex in question is neighbours with a member of the dominating set without considering whether it is part of the set itself.

%The binary bosonic solver is been updated since the original paper and now includes a gradient-free optimisation which requires fewer evaluations of the cost function which should in-turn accelerates the runtime. 
\subsection{Optimisation algorithm pseudocode}
Pseudocode of the binary bosonic solver with gradient-free optimisation is shown in Algorithm~\ref{alg:VQA2}.
%\notess{How can a probability be drawn from $U(-5,5)$??? - Explained}
The parameters $[p_i]$ (line \ref{algline:flip}) are used to define a sigmoid function that then controls the bit flipping probabilities. 
Line \ref{algline:classicalopt} is different from that used in \citet{Park2026-sy} in that we are now using a gradient-free optimisation. 
The default gradient-free optimisation uses the Nevergrad package, but users can specify their own classical optimisation algorithm if desired \citep{Teytaud2018-rp}.
%\notejp{new below} 
For these tests, the default Nevergrad optimisation is used, which has a learning rate parameter $\lambda$.
The other change from previous work is that we here use an all-one input state, $\ket{1} ^{\otimes n}$, rather than the alternating $\ket{01} ^{\otimes \frac{n}{2}}$ state from the previous study. 
%\notess{do you mean `alternating'?- Yes, changed}
This change was recommended to us by ORCA Computing and is considered the default in their updated software development kit.

\algdef{SE}[REPEATN]{RepeatN}{End}[1]{\algorithmicrepeat\ #1 \textbf{times}}{\algorithmicend}

\begin{algorithm}[tp]
\caption{Variational Quantum Algorithm}\label{alg:VQA2}
\textbf{inputs:} cost function $F(\bm{x})$, input state $\ket{1}^ {\otimes n}$, learning rate $\lambda$ \newline
\textbf{outputs:} optimum state $x_{min}$, minimum energy $E_{min}$
\begin{algorithmic}[1]
\State $\bm{\theta} := [\theta_j]$, initialise interferometer parameters randomly from $U(0,\frac{\pi}{2})$
\State $\textbf{p} := [p_i]$, initialise flipping parameters randomly from $U(-5,5)$
\RepeatN{NIter}
    \RepeatN{NSamp}
        \State pass input state through interferometer 
        \State  measure photons in each mode $\ket{out}$
        \State $\bm{x}$ := threshold map $\ket{out}$ to bit string   \label{algline:convert}
        \State $\bm{x}$ := flip or hold bits depending on $\textbf{p}$  \label{algline:flip}
        \State state energy $E := F(\bm{x})$
        \End
    \State $\bm{\theta}$, \textbf{p} := update with Nevergrad optimisation function $f_{ng}(\lambda)$ \label{algline:classicalopt}
    \End
\State \textbf{return} $\bm{x}_{min}$, $E_{min}$
\end{algorithmic}
\end{algorithm}

Parameter values held constant throughout are the number of iterations ($\mbox{NIter}=400$), the number of samples per iterations ($\mbox{NSamp}=600$), and the learning rate used in the classically computed gradient-free optimisation ($\lambda = 1\times 10^{-2}$).
%\notess{where is the learning rate used in the pseudocode? - Addressed in text and psuedocode}

\subsection{Runtime components}
\begin{figure}
    \centering
    \includegraphics[width=1\linewidth]{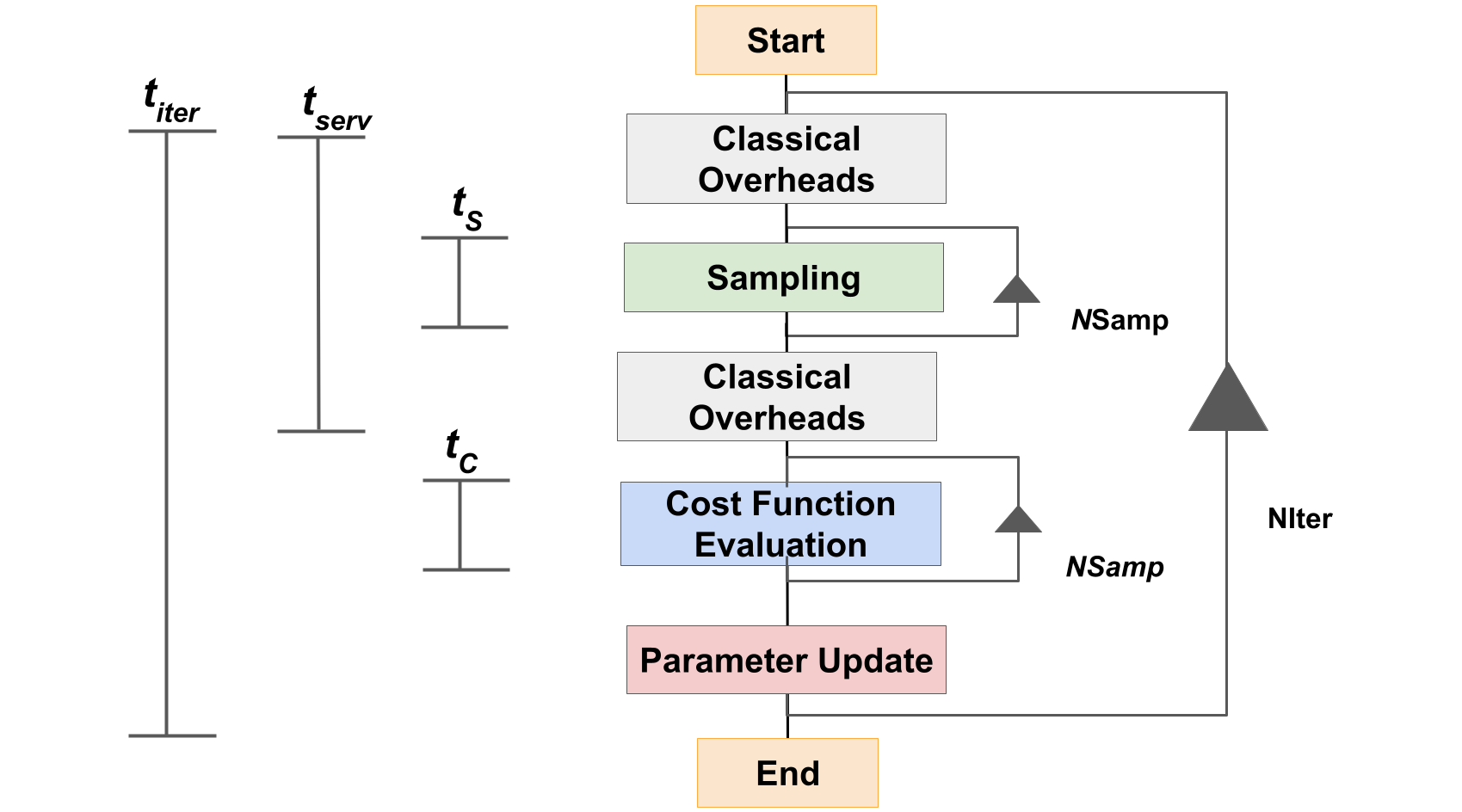}
    \caption{A visual representation of the Binary Bosonic Solver run in practice on the PT-2 device with the various timing metrics labelled. $t_{iter}$ is the time for a single iterations; $t_{serv}$ is the time taken accessing the PT-2 device including classical overheads and sampling; $t_S$ is the time for an individual sample to be generated from the PT-2 device; $t_C$ is the time taken for the local CPU to evaluate the cost function for a single bit string instance.
    %\notess{Change MaxIter, MaxSamp to NIter, NSamp - Changed}
    }
    \label{fig:TimingDiag}
\end{figure}

The algorithm can be visualised as in Figure~\ref{fig:TimingDiag}.
This is helpful when considering the breakdown of the total runtime in terms of the individual component steps and where physically those steps are being executed.

The measures of interest in this study are the size of the dominating set found and the wall-clock runtime.
The wall-clock runtime includes everything from Figure~\ref{fig:TimingDiag}: the sampling from the PT device, the evaluations of the cost function on the sampler outputs, and the classically-computed updates of the interferometer parameters. 
The total runtime can be expressed as:
\begin{equation}
    t_{tot} = \mbox{NIter}(2t_L+\mbox{NSamp}(t_S+t_C)+t_P)
    \label{eq:T}
\end{equation}
where $t_L$ is the classical overheads including internet latency, $t_S$ is the time taken to produce a sample, $t_C$ is the time per cost function evaluation and $t_P$ is the time required to update the parameters.
% \notess{$S$ and $C$ do not appear in the eqn!
% Do you mean $t_{samp}$ and $t_{eval}$?
% If so, use the former to make the eqn shorter -- and update fig2 with these names - Changed}

In practice, the algorithm can be set to terminate when the best found solution has not changed in a given number of iterations, and we say that the training has converged.
Here, however, all runs continue to the maximum number of iterations (NIter).
By recording the current best solution found by each iteration we can then estimate the earliest iteration at which the algorithm could have been terminated without increasing the found set size, $I_{con}$. 
The time taken to get to this iteration, we denote as the convergence time $t_{con}$, which can be calculated as:
%\notess{???}

\begin{equation}
    t_{con} = \frac{t_{tot} \times I_{con}}{\mbox{NIter}}, 
    \label{eq:C}
\end{equation}
where $t_{tot}$ is the total runtime, and $I_{con}$ 
%\notess{do you mean $I_{con}$?}
is the iteration at which the best solution has not improved for a set number of consecutive iterations. For the results presented here, we set that number to be 50. 

As mentioned above, the PT-2 device is capable of running a time-bin interferometer with two loops. To consider the effect of the two-loop set-up, both the single and double loop configurations are run on the real hardware. 
Both sets of results from the ORCA PT-2 sampler (single and double loop) are compared to the classical results from the previous study.  

\section{Results}\label{sec:Results}

\begin{figure}
    \centering
    \includegraphics[width=0.7\linewidth]{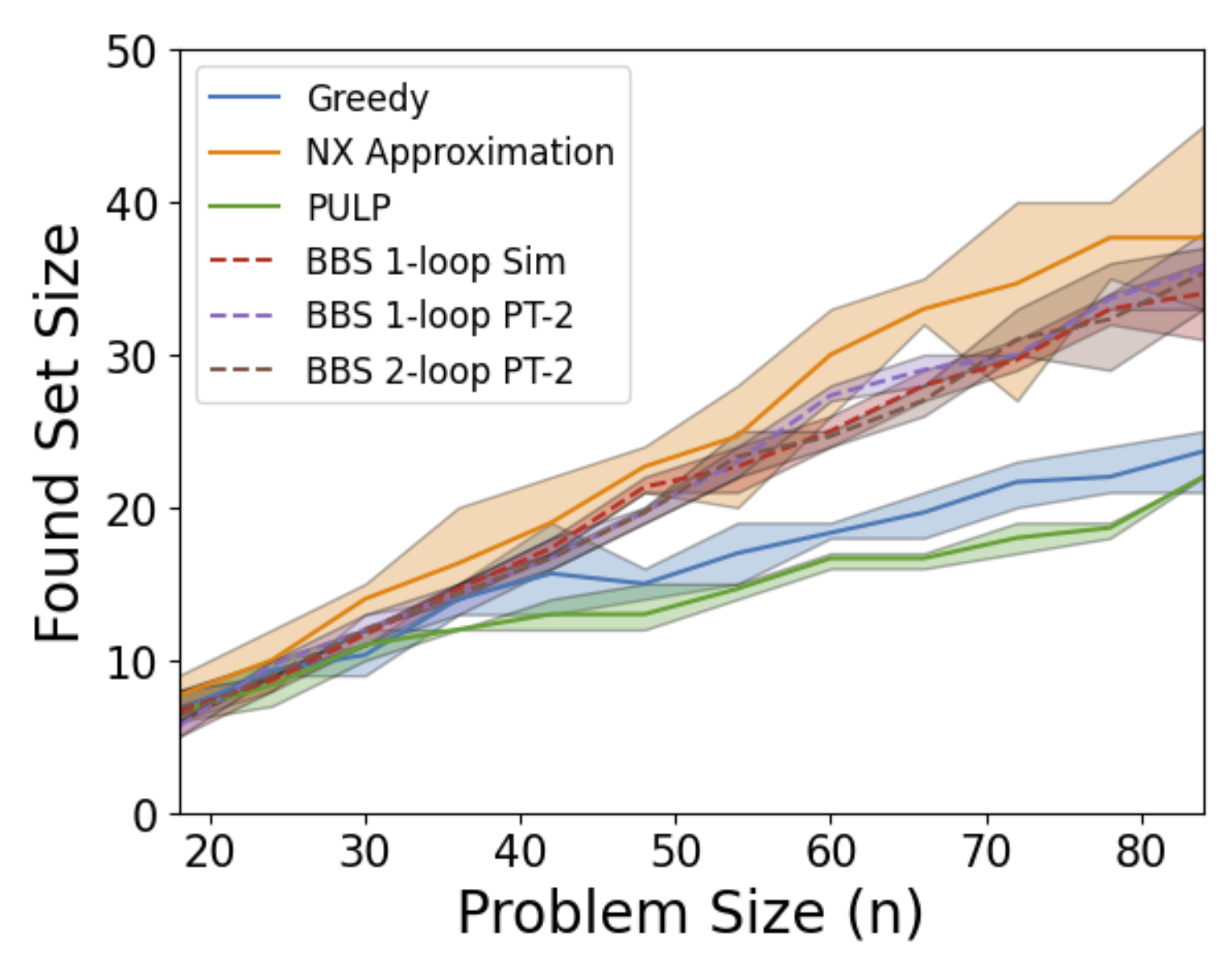}
    \caption{Comparisons between the found solution size of three classical methods for solving the dominating set problem (Greedy, NX Approximation and PULP) and three implementations of the binary bosonic solver (BBS); a 1 loop CPU simulation (BBS 1-loop Sim), a 1-loop PT-2 implementation (BBS 1-loop PT-2), and a 2-loop PT-2 implementation (BBS 2-loop PT-2). The lines represent the mean value over three different graphs (for each value of $n$), and the shaded areas shows the minimum and maximum of results.
    %\notess{3 graphs at each $n$? So three results? or a range of $p$ values too?  Try expressing the expt itself as an algorithm. - Clarified including in method section}
    }
    \label{fig:Comp2}
\end{figure}

\subsection{Solution quality: size of found sets}

In terms of comparing solution quality, we look at the minimum size of the dominating set found by each method. 
Results come from the three classical methods (Greedy, NetworkX approximation, and PULP), and three implementations of the binary bosonic solver (BBS); a 1 loop CPU simulation (BBS 1-loop Sim), a 1-loop PT-2 implementation (BBS 1-loop PT-2), and a 2-loop PT-2 implementation (BBS 2-loop PT-2).  

%\notejp{new below}
The results from the three classical methods are taken from the previous study \citep{Park2026-sy} as they are not affected by the changes to either the PT hardware or the BBS algorithm. 
Due to these changes, all of the BBS results (both in hardware and in simulation) are new to this paper. 
As discussed in the experimental design section, for each graph size, $n$, three random graphs are generated using three different values of the seed $s$. 
These are then fixed for use with both the quantum algorithm implementations and classical comparison. 

%\notess{note here that the classical results are from prev study, as nothing has changed in them -- but all the BBS/PT runs are new, as alg/inputs/hw have changed -- move the earlier discussion about using the same graphs as before to here -- later in results sec, discuss if change of alg/hw changed the results qualitatively/quantitatively from prev paper.}

%the PT-2 device running both a single- and a double-loop configuration, and the CPU simulation of a single-loop time-bin interferometer running the BBS algorithm. 
%\notess{is this CPU-sim the BBS?  Use consistent terminology - Edited, open to discussion}
This is shown in figure~\ref{fig:Comp2}.
It has not been possible to obtain a comparable set of results for the CPU simulation of 2-loop BBS algorithm because the runtimes are increasingly prohibitive. 
This is discussed in more detail below.

At smaller $n$ the BBS algorithm finds comparable set sizes to the classical algorithms in all implementations. 
At the larger tested $n$, performance of the BBS algorithm drops off.
We hypothesise that this is because not enough of the search space is being explored with the chosen numbers of samples and iterations. 
This hypothesis could be tested in future work, but in practical applications, where use of the quantum device will likely be paid for either by runtime or by total number of samples, there may be restrictions on choice of NSamp and NIter. 
Being able to estimate a sufficient number of samples and iterations for a required confidence in the answer would be a beneficial study to conduct in the future.

%\notejp{New}
The similarity between the BBS implemented on the 1-loop software simulator and the 1-loop PT-2 hardware helps to validate confidence in the simulation process, which may help in future work with limited access to the real device(s). 
There does not seem to be a discernable difference between the the performance of the one loop and two loop PT-2 implementations across the range of problem sizes tested here, even where there are smaller dominating sets available to find (as seen by the set sizes found by the classical Greedy and PULP algorithms). 
The shaded areas that show the minimum and maximum sizes of the sets found are reasonably narrow, which could again suggest that the the number of samples and iterations are not adequate to search for particularly good areas of the space.

%\notess{More discussion of Fig3 -- esp the PT results!!}

The PULP linear programming approach should provide exact solutions (and therefore a fixed benchmark) to the problem.
However, in practice there are approximation errors. 
The problem is optimised using real numbers rather than integers, which means that when applied to an combinatorial optimisation problem such as this, some rounding is applied to the end result. 
This rounding can cause a suboptimal result to be output. 
This is seen in Figure \ref{fig:Comp2} where the Greedy algorithm actually finds a smaller dominating set than PULP for one of the 30 node instances.

%\notejp{new below} 
The results in terms of found set size are very similar to those seen in the previous study that looked only at CPU simulations: the BBS implementations find dominating sets that are smaller than the NX Approximation but larger than the other two classical methods. 
%\notess{Erm -- I though 2-loop was prohibitive to simulate?}
To effectively compare the results of this paper to the previous study it would be useful to better understand the effect of the maximum number of samples and iterations used in the BBS algorithm. These determine the number of cost function evaluations used overall and is something discussed in more detail in the remainder of this work.

\subsection{Search space coverage}
Unlike gradient-based optimisation methods, the gradient-free optimisation within the binary bosonic solver (BBS) algorithm does not aim to reduce the value of the cost function after every iteration, therefore this is not the best metric to track the algorithm's progress in this case.
%\notess{what is loss? cost function?? - Edited}
Instead, gradient-free optimisation aims to explore the search space, and therefore it is informative to consider the number of different bit strings (candidate solutions) that are evaluated by the cost function. 
%Unique in this case refers to the binary string that is evaluated, it is highly unlikely that the results of the cost function will be unique in each case.

\begin{figure}
    \centering
    \includegraphics[width=0.7\linewidth]{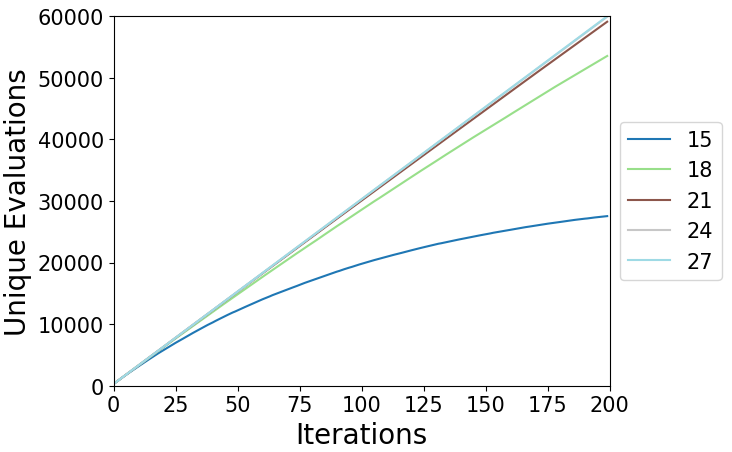}
    \caption{The cumulative number of unique cost function evaluations carried out as the Binary Bosonic Solver progresses through the iterations, for different problem sizes.}
    \label{fig:CostEvals}
\end{figure}

The number of bit strings required to cover the entire space is  $2^n$; the maximum number of different cost function evaluations in the BBS solver is NSamp $\times$ NIter, which in this experiment is $240,000$
(although there is no guarantee that there are no repeated bit string candidate solutions). 
Therefore only problems with $n\leq17$ can be fully sampled,  equivalent to exhaustive search.
%\notess{why do you say the bit in parens? the problem is hard, implying worst case you might need to explore every config. - Because otherwise there's no point doing it if we expect to need exhaustive search in every case? - Edited}
This can be seen in Figure \ref{fig:CostEvals}, where the $n=15$ line levels out at higher iterations.
For $n=18$ and $n=21$, there is some deviation from the linear trend which suggests that there are some repeated candidate bit strings in these cases. The BBS algorithm caches previous evaluations, so the cost function is not recalculated when a previously seen bit string is generated.
In all of the  other, larger experiments, the number of unique cost evaluations continue to scale linearly, which implies that the BBS is succeeding in exploring new areas of the search space with each iteration. 

%It should be noted that full exploration of the search space is not required to reach close to optimal solutions. 
Gradient-based optimisation is sensitive to initial parameter choice; 
therefore there may be a case for beginning a large optimisation problem with a gradient-free approach and later using the gradient-based version to fine-tune a close-to-optimal solution. 
Such a hybrid BBS approach would be an interesting and potentially  beneficial avenue for future work. 

Another benefit of the gradient-free approach is that it may find multiple distinct regions of the search space that contain near-optimal solutions. 
This may be particularly useful in certain use cases where a robust set of solutions is required, rather than a single result.
This could allow users to down-select based on additional factors or constraints that could not be formalised in the cost function or that are unknown at the time of running the algorithm.

\subsection{Run times}

\begin{figure}
    \centering
    \includegraphics[width=1\linewidth]{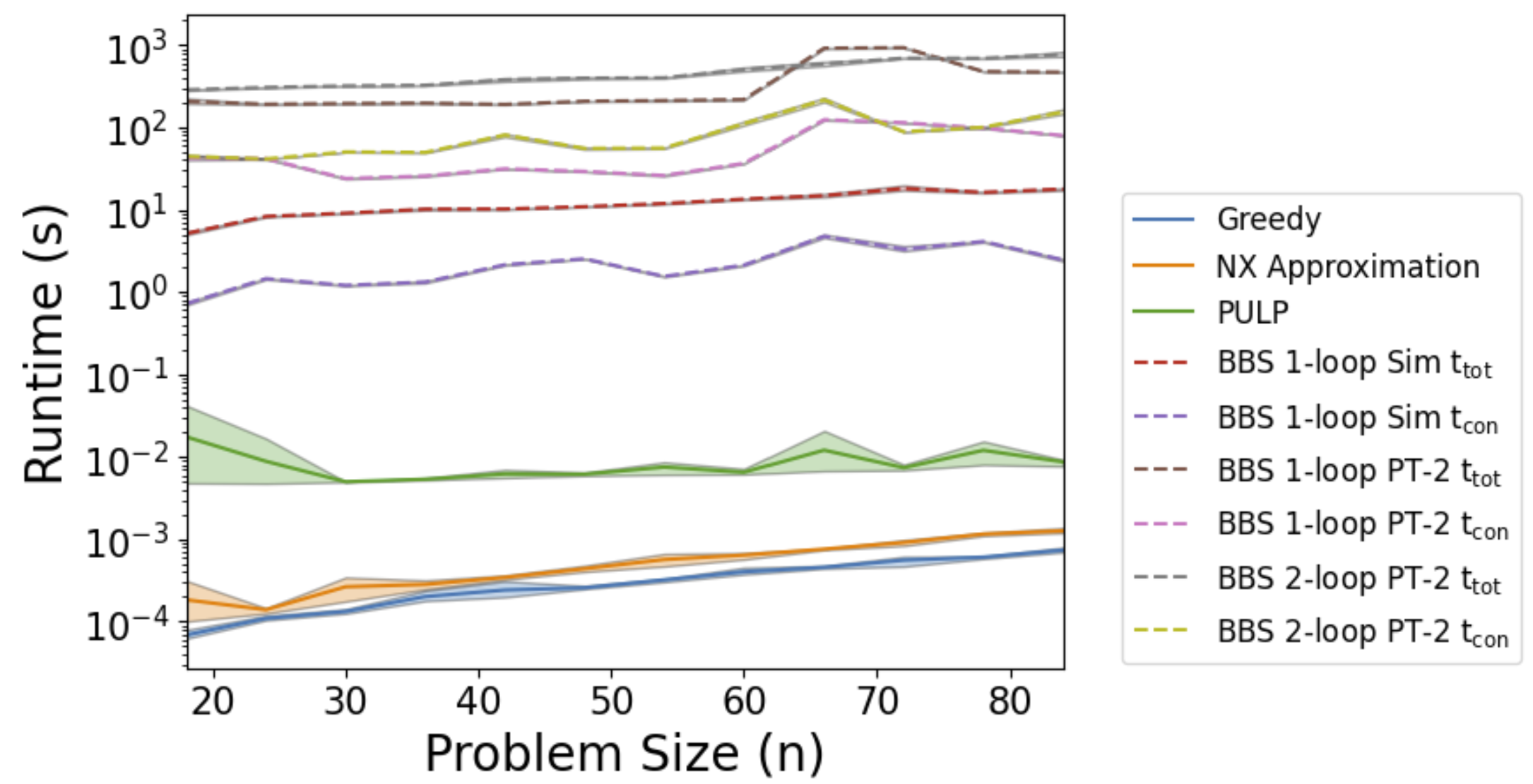}
    \caption{Comparisons between the runtime of three classical methods for solving the dominating set problem (Greedy, NX Approximation and PULP) and three implementations of the binary bosonic solver (BBS); a 1 loop CPU simulation (BBS 1-loop Sim), a 1-loop PT-2 implementation (BBS 1-loop PT-2), and a 2-loop PT-2 implementation (BBS 2-loop PT-2). For the BBS implementations, both the total runtime $t_{tot}$ and the estimated convergence time $t_{con}$ are shown with convergence time, $t_{con}$, calculated by equation \ref{eq:C}. Note the log scaling on the y-axis. The lines represent the mean value over three different graphs (of the same $n$ and $p$ values) and the shaded areas shows the minimum and maximum values of results.}
    \label{fig:Comp1}
\end{figure}

Figure \ref{fig:Comp1} shows the wall-clock runtime for the three classical algorithms, and the three BBS implementations (1-loop Sim, 1-loop PT-2 and 2-loop PT-2). 
%of the PT-2 and the BBS algorithm as run on the real PT-2 device. 
%\notess{Okay, so BBS is not the same as the sim -- need to make clear exactly what is meant be these terms, early on. - Edited for consistency with above}
There are also lines that show the estimated convergence time as calculated by \ref{eq:C} for both the real and simulated PT-2 device running the gradient-free BBS algorithm. 
Overall the runtime for the BBS algorithm is higher than that of the classical methods whether being enacted on a CPU simulation or with the real PT-2 device. 
%\notejp{new below} 
Although higher than the classical methods, the runtime of the BBS simulations in this experiment are approximately an order of magnitude smaller than shown in the previous study \citep{Park2026-sy}.
This may be due to the fewer cost function evaluations that are required per iteration of the gradient-free BBS algorithm compared to the previous gradient-based version.
As in the previous study the runtime does not appear to scale with $n$.
These further results add weight to the hypothesis formed from the previous results, that the scaling of BBS runtime is not dominated by the graph size and that this implementation could be a beneficial method for large problems if the hardware can be scaled appropriately. 

It should also be noted that despite not being shown in the data collected from this experiment, the previous study showed that beyond a problem size of around $n=100$, the PULP algorithm begins to scale exponentially in time.

The PT-2 runs (real device and simulated)
%\notess{again. no 2-loop sims?}
include the same cost function calculations and parameter update steps run on the same CPU. This means, according to equation \ref{eq:T}, that the difference in runtime must be due to the physical sampling. 
We have not been able to measure these quantities directly in this work, however ORCA Computing report that the combined classical overheads (including internet latency) and sampling time tends to be $\approx$\,300\,ms, independent of the number of samples requested (within the tested range 1--1000).
This suggests that the the actual time taken to produce the samples is orders of magnitude shorter than the classical overheads. 

\begin{figure}
    \centering
    \includegraphics[width=0.7\linewidth]{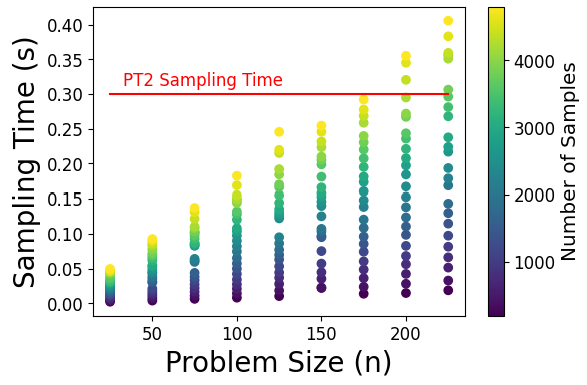}
    \caption{Time taken to draw different numbers of samples from a simulated single loop time-bin interferometer for a range of problem sizes. The red line denotes the approximate time required to complete the equivalent step using the PT-2 device.}
    \label{fig:SampTime}
\end{figure}

Figure \ref{fig:SampTime} shows how time required to produce samples using the state-vector simulator of a single loop time-bin interferometer scales with both the number of samples requested and the size of the input state $n$ (which for this study has always been equal to the size of the problem). 
In the CPU simulation of the boson sampler, the sampling element happens via a \enquote{chain-rule} calculation that works out the probabilities of seeing different photon numbers in the first mode of the interferometer, samples from this distribution, then uses this to calculate the conditional probability on the second mode, and so on.
This calculation is repeated for every sample requested by the user.
Further details of the simulation method can be found in \citep{Bulmer2022-bw}.
From figure~\ref{fig:SampTime}, it is clear that drawing large numbers of samples linearly increases the runtime, which is expected from this method of simulation.
For the range of problem sizes used in the real experiment ($n<90$), the simulated sampling times are below the combined classical overheads and sampling times quoted from ORCA for the real device ($\approx$\,300\,ms; shown as the horizontal line in Figure \ref{fig:SampTime}).
Although the entire BBS algorithm was tested only for problem sizes $n<90$, it was possible to measure the sampling time required for the CPU simulation for $n<225$ and $\mbox{NSamp}<5000$. 
Plotting these shows that, % by plotting a larger range of problem sizes and samples, we can see 
there is a point at which the PT-2 device should produce (and deliver) samples faster than the CPU simulation. This may be the point at which quantum advantage can be seen. 
%\notess{not clear where these larger $n$ value results have come from -- did you so these measurements? - Added lines above}

It is theoretically possible to simulate a full state-vector calculation of the interferometer only the once, find the complete probability distribution, then draw an unlimited number of samples at little extra cost. 
However, this initial calculation is what makes boson sampling a \#P  problem, and quickly becomes intractable for classical computing at increasing problem sizes.
For moderate problem sizes, and a large number of samples, this full distribution calculation would likely be more performant than the current simulation.
For larger problem sizes, chain-rule methods allow a better scaling in problem size at the cost of a worse scaling in sample number.

\begin{figure}
    \centering
    \includegraphics[width=0.7\linewidth]{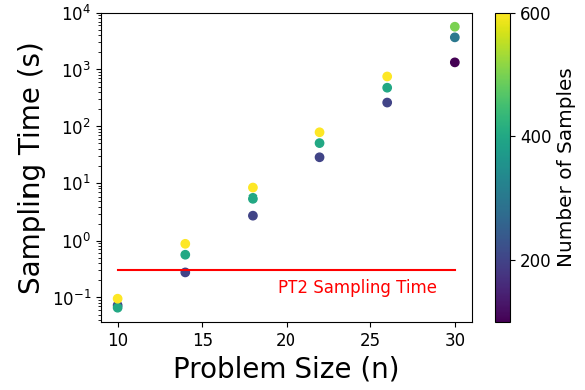}
    \caption{Time taken to simulate a double loop time-bin interferometer for a range of problem sizes and draw range of samples. Note the log scale on the y-axis when comparing to Figure \ref{fig:SampTime}.}
    \label{fig:SampTime2}
\end{figure}

As mentioned previously, the simulation of a double-loop time-bin interferometer has run-time prohibitive for producing a full set of comparable data. 
From the structure of the algorithm, we know that it must be the sampling time that increases dramatically with the double-loop simulation. 
This is shown in Figure \ref{fig:SampTime2} over a range of small problems. 

\begin{figure}
    \centering
    \includegraphics[width=0.7\linewidth]{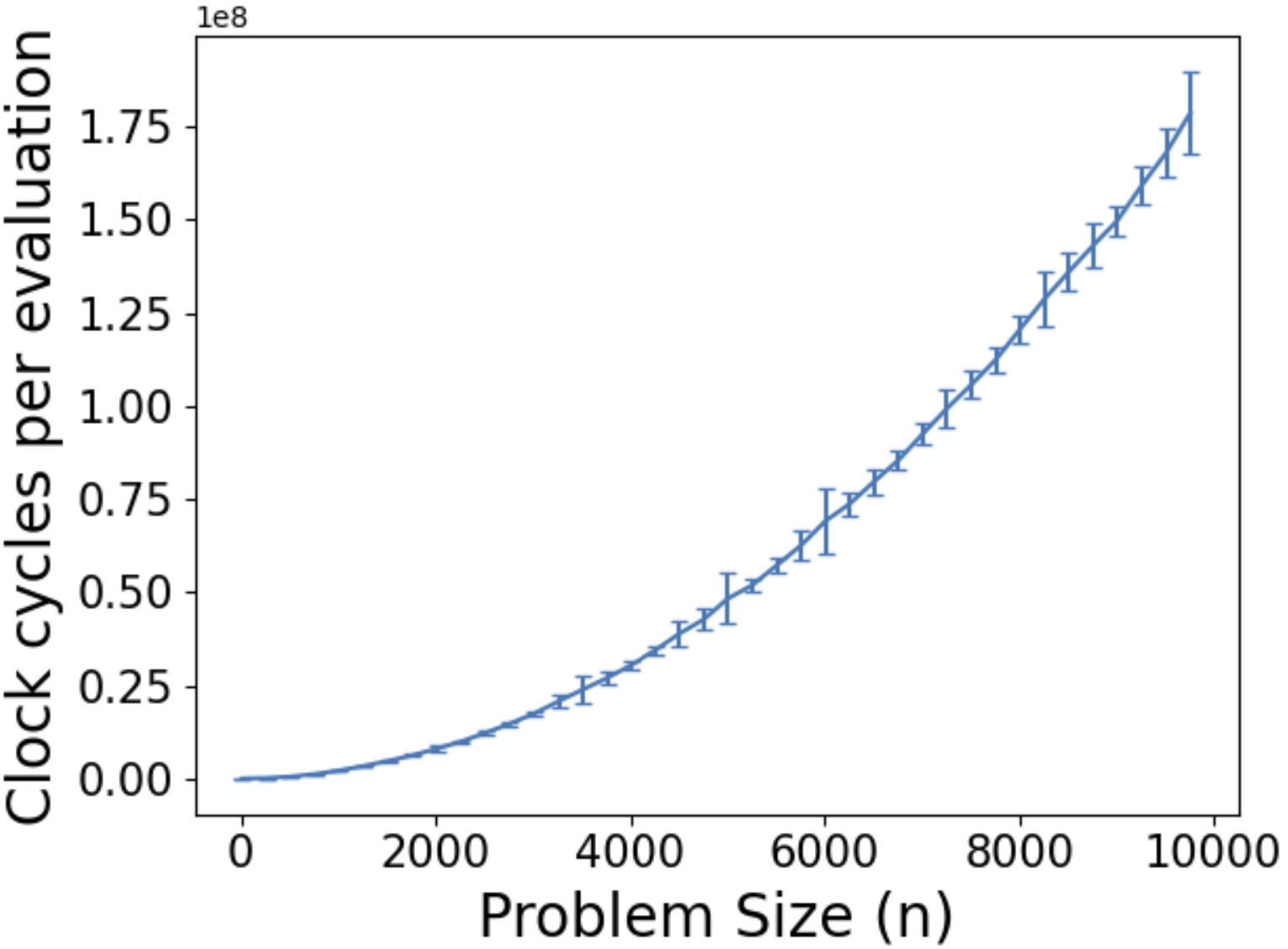}
    \caption{The number of CPU clock cycles taken to calculate a single instance of the cost function against the problem size. The problem size represents the number of nodes in the graph. All graphs were produced with a connection density $p=0.05$. For each instance, 10 unique calculations are performed. The plot shows the mean and standard deviation across these calculations.}
    \label{fig:CostF}
\end{figure}

As seen in equation \ref{eq:T} and Figure \ref{fig:TimingDiag}, the cost function needs to be evaluated once for every sample within every iteration.
If this is a costly (in time) calculation (if $t_{C}$ if large compared to $t_{S}$), it will have a great impact on the total runtime.
Therefore it is worth better understanding how this scales with $n$. 
%\notejp{Added below to justify measuring clock cycles instead of wallclock time}
Wall-clock runtime varies greatly depending on the background processes being run on the device and therefore, we have chosen to measure the clock cycles required for a single cost function evaluation for a range of problem sizes. 
This is shown in Figure \ref{fig:CostF}.
Wall-clock time is assumed to follow the same scaling but with greater variability and magnitudes based on the capability of the CPU.
From this result, it is clear to see that simplifying and optimising the cost function should be an area of focus in any practical use of the BBS algorithm, especially when considering large problems.
Calculating the cost function for each sample within a single iteration of the BBS algorithm is an entirely parallelisable problem. 
Once all of the samples have been obtained from the device, each cost function evaluation can happen independently, which provides the potential for  speed up depending on the quantity of compute resources allocated to the task.

%\notejp{Will only end up including one of the cost function plots}
%\begin{figure}
%    \centering
%    \includegraphics[width=1\linewidth]{Results/CostFLog.png}
%    \caption{ALTERNATIVE FIGURE 8 (LOG SCALING) Time to calculate a single instance of the cost function against the problem size. The problem size represents the number of nodes in the graph. All graphs were produced with a connection density $p=0.05$. For each instance, 10 unique calculations are performed. The plot shows the mean and standard deviation. Note the log scaling on this plot.}
%    \label{fig:CostF-alt}
%\end{figure}

Given the total number of evaluations of the cost function for a complete run of the BBS algorithm is NSamp $\times$ NIter (assuming all unique bit strings), it would be worth a deeper study into the balance between the chosen value of NSamp and NIter. 
A large enough number of samples is required to \enquote{well map} the area of the search space that the current parameters provide, and to remain robust to noise and errors in the system. 
A large enough number of iterations allows for more parameter combinations to be tested, thus providing a broader exploration of the search space. 
Exploring this trade-off would be a valuable avenue of future work.

\section{Conclusions and Future Work}\label{sec:Conc}
At small problem sizes, the ORCA PT-2 device running the gradient-free Binary Bosonic Solver (BBS) finds dominating sets of equivalent size to classical methods. 
Within the range explored, at larger problem sizes, higher $n$, the performance of the PT-2 (and simulator) degrades compared to the classical methods.  
We hypothesise that the number of samples and/or iterations is currently restricting performance of the BBS at high $n$, and further work should look to exploring the effects of changing both the number of samples and number of iterations, and the possible trade-off between them.

With respect to the runtime of the whole algorithm, the largest factors in determining the runtime (using the PT-2 device) are the  number of samples, the number of iterations, and the calculation of the cost function. 
As mentioned above, the number of samples and iterations should both be made as large as possible to allow the greatest possible exploration of the search space.
Therefore, the most promising ways to optimise the runtime would be (a) to determine the minimum viable number of samples and iterations to achieve the required solution quality and/or allow the algorithm to terminate at a required solution quality, 
and (b) to optimise the design and parallelisation of the cost function calculations.

It would also be beneficial to consider the use of a hybrid algorithm of gradient-free initial \enquote{wide area} search and then gradient-based fine tuning to a better solution. 
%\notess{Still no guarantees of optimality -- just fine tune to get a better solution - Changed}
This may allow a user to benefit from the computational efficiency of the gradient-free approach whilst maximising the likelihood of finding a minimal solution.

%\backmatter

\section*{Declarations}

\subsubsection*{Ethical Approval} Not applicable, no human or animal research involved.

\subsubsection*{Competing Interests} The authors have no competing interests of a financial or personal nature, nor other interests that might be perceived to influence the results and/or discussion reported in this paper.

\subsection*{Authors' contributions} J.P. wrote the first draft of the manuscript and performed all simulation experiments. S.S. and I.D'A. reviewed the manuscript, and contributed to the design of the experiments and analysis of results. 

\subsubsection*{Funding} The authors wish to acknowledge Defence Science Technical Laboratory (Dstl) who are funding this research. 
Content includes material subject to © Crown copyright (2026), Dstl. This material is licensed under the terms of the Open Government Licence except where otherwise stated. To view this licence, visit http://www.nationalarchives.gov.uk/doc/open-government-licence/version/3 or write to the Information Policy Team, The National Archives, Kew, London TW9 4DU, or email: psi@nationalarchives.gov.uk

\subsubsection*{Data Availability}
Supporting information is available from the authors.

\subsubsection*{Acknowledgments}

Access to a PT-2 machine has been facilitated by ORCA Computing. ORCA Computing have also been very helpful in the discussion and understanding of the hardware details.

% \bigskip


% Springer journals and proceedings: \url{https://www.springer.com/gp/editorial-policies}

% Nature Portfolio journals: \url{https://www.nature.com/nature-research/editorial-policies}

% \textit{Scientific Reports}: \url{https://www.nature.com/srep/journal-policies/editorial-policies}

% BMC journals: \url{https://www.biomedcentral.com/getpublished/editorial-policies}
% \end{flushleft}

\begin{appendices}

%\section{Section title of first appendix}\label{secA1}

%An appendix contains supplementary information that is not an essential part of the text itself but which may be helpful in providing a more comprehensive understanding of the research problem or it is information that is too cumbersome to be included in the body of the paper.




\end{appendices}


%\bibliographystyle{apalike}
\bibliography{paperpile}% common bib file

\end{document}